\documentclass[aps,prx,reprint,onecolumn,superscriptaddress]{revtex4-2}
\usepackage{amsfonts,amssymb,amscd,amsthm,amsmath,graphicx,bm,mathrsfs,physics,units,bbm,braket,tabularx,dcolumn,graphicx,times,booktabs,multirow}
\usepackage[colorlinks=true]{hyperref}
\hypersetup{linkcolor=magenta, urlcolor=blue, citecolor=blue,pdfstartview={FitH}, unicode=true}
\usepackage[linesnumbered,ruled,vlined]{algorithm2e}
\usepackage{makecell}
\begin{document}

\title{Quantum Capsule Networks}

\affiliation{Center for Quantum Information, IIIS, Tsinghua University, Beijing 100084, P. R. China}
\affiliation{Shanghai Qi Zhi Institute, 41th Floor, AI Tower, No. 701 Yunjin Road, Xuhui District, Shanghai 200232, China}
\affiliation{International Research Centre MagTop, Institute of Physics, Polish Academy of Sciences, Aleja Lotnikow 32/46, PL-02668 Warsaw, Poland}

\author{Zidu Liu}\thanks{These authors contributed equally to this work.}
\affiliation{Center for Quantum Information, IIIS, Tsinghua University, Beijing 100084, P. R. China}

\author{Pei-Xin Shen}\thanks{These authors contributed equally to this work.}
\affiliation{Center for Quantum Information, IIIS, Tsinghua University, Beijing 100084, P. R. China}
\affiliation{International Research Centre MagTop, Institute of Physics, Polish Academy of Sciences, Aleja Lotnikow 32/46, PL-02668 Warsaw, Poland}

\author{Weikang Li}
\affiliation{Center for Quantum Information, IIIS, Tsinghua University, Beijing 100084, P. R. China}

\author{L.-M. Duan}\email{lmduan@tsinghua.edu.cn}
\affiliation{Center for Quantum Information, IIIS, Tsinghua University, Beijing 100084, P. R. China}
\author{Dong-Ling Deng}\email{dldeng@tsinghua.edu.cn}
\affiliation{Center for Quantum Information, IIIS, Tsinghua University, Beijing 100084, P. R. China}
\affiliation{Shanghai Qi Zhi Institute, 41th Floor, AI Tower, No. 701 Yunjin Road, Xuhui District, Shanghai 200232, China}
\date{\today}

\begin{abstract}
    Capsule networks, which incorporate the paradigms of connectionism and symbolism, have brought fresh insights into artificial intelligence. The capsule, as the building block of capsule networks, is a group of neurons represented by a vector to encode different features of an entity. The information is extracted hierarchically through capsule layers via routing algorithms.
	Here, we introduce a quantum capsule network (dubbed QCapsNet) together with an efficient quantum dynamic routing algorithm. To benchmark the performance of the QCapsNet, we carry out extensive numerical simulations on the classification of handwritten digits and symmetry-protected topological phases, and show that the QCapsNet can achieve an enhanced accuracy and outperform conventional quantum classifiers evidently. 
    We further unpack the output capsule state and find that a particular subspace may correspond to a human-understandable feature of the input data, which indicates the potential explainability of such networks. Our work reveals an intriguing prospect of quantum capsule networks in quantum machine learning, which may provide a valuable guide towards explainable quantum artificial intelligence.
\end{abstract}
\maketitle

\section{Introduction}
Connectionism and symbolism are two complementary approaches towards artificial intelligence (AI) \cite{Russell2020Artificial}.
Inspired by biological brains, connectionism aims to model the intelligence as an emergent phenomenon by connecting a large number of neurons. The most popular paradigm of this approach lies on artificial neural networks. With the recent rise of deep learning \cite{Lecun2015Deep,Jordan2015Machine,Goodfellow2016Deep,Bibbo2022Neural}, connectionism has attracted great attentions and become one of the most promising ways to realize general AI. Noteworthy examples \cite{Goodfellow2016Deep} include feed-forward neural networks, recursive neural networks, recurrent neural networks, and convolutional neural networks (CNNs). Connectionism AI requires less prior knowledge, which is beneficial for scaling up and broader applications. Yet, connectionism AI typically treats the machine learning model as a black box, 
which makes it challenging for human to explain the rational behind the model's decisions.
In comparison, symbolic AI is based on the logic, deduction, and higher-level symbolic representations \cite{Gilmore1960Proof, Eliasmith2006Symbolic}. It enables one to trace back how the decision is made, and seek for the interpretability of machine learning models. However, such models generally require more sophisticated prior knowledge, which hinders the widespread application of symbolic AI.

\begin{figure}[htbp]
	\includegraphics[width=0.6\textwidth]{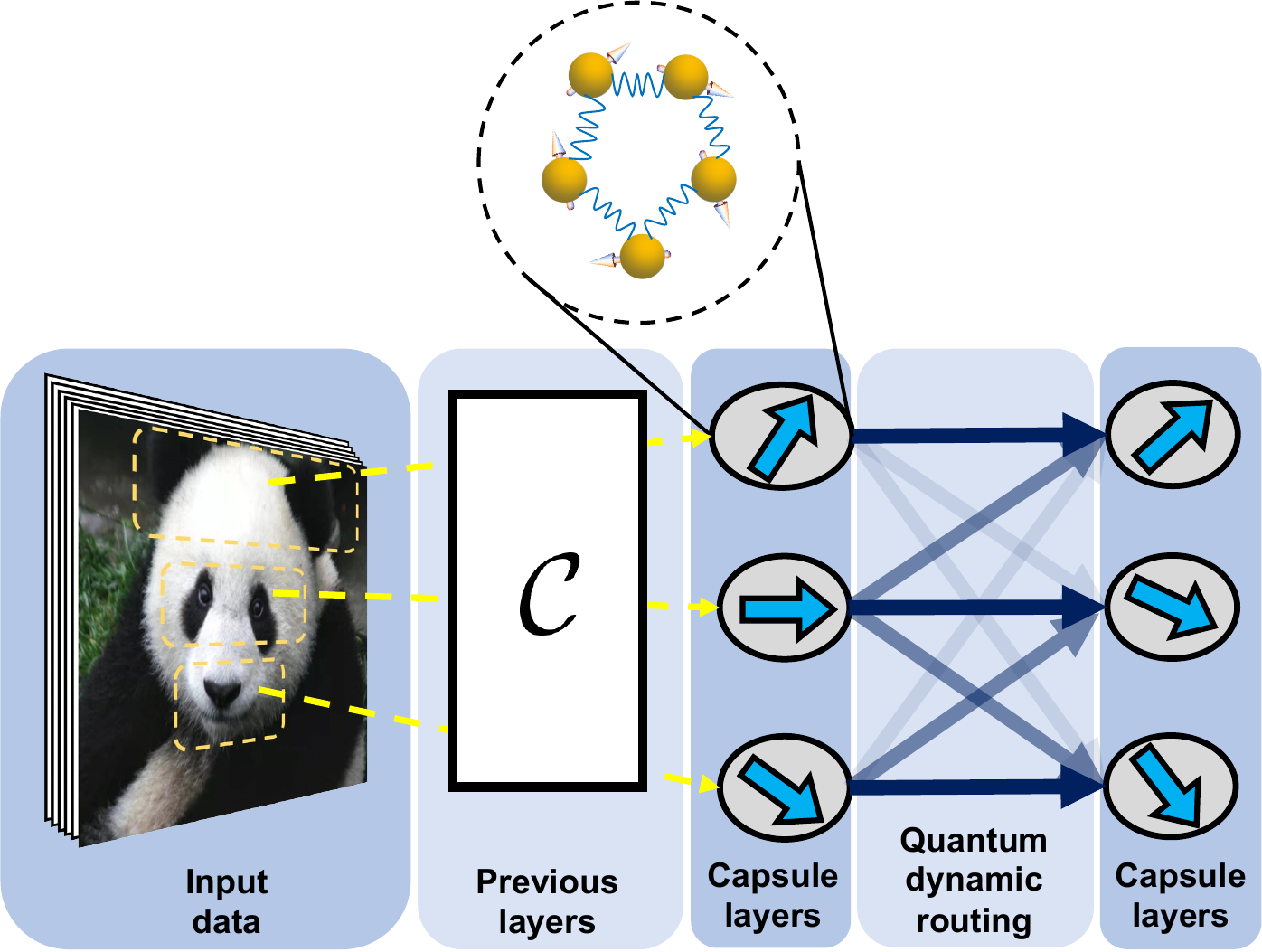}
	\caption{General framework of quantum capsule networks (QCapsNets). The input data is first preprocessed in the previous layers to extract some preliminary features. Within QCapsNets, there are multiple capsules in each layer.
	Each capsule contains a group of interacting qubits, which forms a sub-quantum neural network (sub-QNN). The output states of the sub-QNN are represented by quantum states living in the Hilbert space, which encode features of the entities.  Different capsules can represent different entities, such as ears, eyes and mouth of the panda. The information is processed layer by layer via a quantum dynamic routing algorithm. During the routing procedure, the probability that the lower-level capsule are assigned to the higher-level one will be updated with their geometric agreements.
	As such, higher-level capsules can not only recognize the active entities in the lower layer, but also preserve their geometric relationships. The magnitude of the output capsule indicates the classification probability. This Giant Panda image has been obtained by the author(s) from the Wikimedia website, where it is stated to have been released into the public domain. It is included within this article on that basis.}
	\label{fig:general_framework}
\end{figure}

Recently, a new paradigm that incorporates both symbolism and connectionism has become an encouraging trend for AI. Many attempts have been made \cite{Hu2014SparsityRegularized,Shi2017ZhuSuan,Dong2020Geometric,Hinton2011Transforming}, among which the capsule network (CapsNet) \cite{Hinton2011Transforming,Sabour2017Dynamic,Hinton2018Matrix,Wang2018Optimization,Patrick2019Capsule}, 
as a variant of the traditional neural network,
has become a cutting-edge research area. In analogy to a group of neurons constituting a specific functional area of the human brain, the basic unit of CapsNets---capsule---is a set of neurons that used to detect a specific entity. Such a capsule is represented by a vector, rather than a scalar in traditional neural networks.
In this fashion, a capsule is able to represent different features of an entity, such as pose, hue, texture, deformation, etc. Furthermore, the norm of the capsule indicates the likelihood that an entity being captured. During the feed-forward process, the information can be transferred layer by layer via the routing-by-agreement mechanism \cite{Sabour2017Dynamic}: 
the capsule in the higher-level layer is predicated upon its geometric relationship (e.g., dot product in the Euclidean space) with the lower-level one. Unlike the max-pooling method throwing away information about the precise position of entities in CNNs, this routing mechanism used in CapsNets can preserve the geometric relationships amongst entities. With such geometric information encoded inside the capsule, CapsNets generally have more intrinsic explainability than CNNs \cite{Shahroudnejad2018Improved, Wang2020Interpretable}.

Quantum computing is based on the computation model following the principles of quantum theory, and is able to harness the features from the quantum world such as superposition and entanglement. The basic memory units of quantum computing are quantum bits, which is also referred to as qubits. In general, for a classical register consisting of $n$ bits,
the state can be described by a length-$2^n$ vector
with $1$ for one entry and $0$ for all the other entries.
For a quantum register consisting of $n$ qubits,
quantum mechanics allows a quantum state to be described by a $2^n$-dimensional complex vector,
which is crucially different from a single non-zero entry in the classical case and usually called the quantum superposition.
With the development and flourishing of quantum computing architecture \cite{Arute2019Quantum, Song2019Generation,Wright2019Benchmarking,Gong2021Quantum,Zhong2020quantum,Wu2021Strong,Zhong2021PhaseProgrammable}, the interplay between quantum physics and AI has attracted a wide range of interests \cite{Biamonte2017Quantum,Dunjko2018Machine,DasSarma2019Machine,Li2021Recent}. Along this line, many heuristic quantum machine learning models have been proposed, including the quantum decision tree classifiers \cite{Lu2014quantum,Heese2022Representation}, quantum support vector machines \cite{Rebentrost2014Quantum}, quantum Boltzmann machines \cite{Amin2018Quantum}, quantum generative models \cite{Lloyd2018Quantum,Dallaire-Demers2018Quantum,Gao2018Quantum,Hu2019Quantum}, quantum convolutional neural networks \cite{Cong2019Quantum,Li2020Quantuma,Kerenidis2019Quantum,Liu2021Hybrid,Wei2022Quantum}, and perception-based quantum neural networks \cite{Beer2020Training}, etc. Some of these works show potential quantum advantages over their classical counterparts, which have boosted the development of quantum AI \cite{Dunjko2018Machine}. Although some recent efforts have been initiated to interpret behavior of AI in quantum optical experiments \cite{Krenn2021Conceptual}, a generally explainable quantum AI is still in its infancy.

Here, inspired by the innovative design of classical CapsNets and the exciting progress in quantum machine learning, we propose a quantum capsule network (QCapsNet), where interacting qubits are encapsulated into a capsule as the building block of the architecture.  For the feed-forward process between the two adjacent layers, we propose an efficient quantum dynamic routing algorithm based on the quantum random access memory. We benchmark our model through the classification tasks of both the quantum and classical data, and show that the capsule architecture can increase the accuracy for the quantum classifiers. In addition, the combination of symbolism and  connectionism prevents the QCapsNet from being fully a black-box model. Indeed, in our QCapsNet, a certain subspace of the capsule encodes explainable feature of the input data. We exam it by tweaking the active output capsule, and show that the one can visualize the variation of a specific explainable feature and semantically interpret the model's output. Our results not only demonstrate an enhanced performance of the QCapsNet in classification tasks, but also reveal its potential explainability, which may pave a way to explore explainable quantum AI. For future applications, QCapsNet may utilized to handle the tasks of learning from classical and quantum data in an explainable way, such as assisting the medical diagnosis, recognizing quantum phase transitions \cite{Ren2022Experimental}, etc.

\section{General framework}

\subsection{Classical capsule networks}

To begin with, we first briefly recap the essential idea of the CapsNet. 
In computer graphics, a set of instantiation parameters is an abstract representation of an entity that is fed into a rendering program so as to generate an image \cite{Hughes2013Computer}.
Here, the entity refers to a segment of an object. For example, nose and mouth are two entities on the face. The motivation behind introducing a capsule is to encapsulate the instantiation parameters as a vector, and thus predict the presence of a specific entity. In this respect, the norm of the capsule vector represents the likelihood that an entity being detected, whereas the components of the capsule vector encode its corresponding features (instantiation parameters), such as pose, deformation, hue, etc. From this perspective, the classical CapsNet is intuitively a neural network that attempts to perform inverse graphics. 

The preprocessing layers of the classical CapsNets (e.g., the convolutional layer) first extract some basic features of the input data, and then encapsulate these features into several capsules to represent different entities. Given the $i$-th capsule vector $\mathbf{v}^l_i$ in the $l$-th layer, the prediction vector $\mathbf{u}^{l+1}_{j|i}$ of the $j$-th capsule in the $(l+1)$-th layer is calculated by $\mathbf{u}^{l+1}_{j|i} = \mathbf{W}_{ij} \mathbf{v}^l_{i}$, where $\mathbf{W}_{ij}$ is a weight matrix to be trained by the gradient descent algorithm. 
After taking into account routing coefficients $r_{ij}$ and a nonlinear normalization function $\texttt{squash}$  {[Eq.~\eqref{eqn:squash}]}, the $j$-th capsule vector in the $(l+1)$-th layer is calculated by $\mathbf{v}^{l+1}_j = \mathtt{squash}(\sum\nolimits_{i} r_{ij} \mathbf{u}_{j|i}^{l+1})$.
The routing-by-agreement algorithm is executed by dynamically updating the routing coefficients $r_{ij}$ with the geometric agreement (such as dot product) $\mathbf{u}_{j|i}^{l+1} \cdot \mathbf{v}_j^{l+1}$. Here, $r_{ij}$ reflects the probability that the lower-level capsule is assigned to the higher-level one. After a few iterations, lower-level capsules with strong agreements will dominate the contribution to the $\mathbf{v}^{l+1}_j$ capsule. The classification result can be deduced from the norm of the output capsule vector (activation probability).  {We present the explicit expression of the $\texttt{squash}$ function and the routing details in Appendix~\ref{sec:Appx_CapsNets} }. By virtue of this algorithm, CapsNets can not only extract information from lower-level capsules, but also preserve their geometric relationships \cite{Sabour2017Dynamic}. As a result, CapsNets can address the so-called Picasso problem in image recognition, e.g., an image with an upside-down position of nose and mouth will not be recognized as a human face in CapsNets. In contrast, CNNs will still classify such an image as a human face, as the max-pooling method generally neglects the spatial relationship between entities \cite{Patrick2019Capsule}.

\subsection{Quantum capsule networks}

We now extend the classical CapsNets to the quantum domain and introduce a QCapsNet model. The general framework of a QCapsNet is illustrated in Fig.~\ref{fig:general_framework}.  For readers unfamiliar with the capsule architecture, we present comparisons among the quantum capsule networks, classical capsule networks, and classical neural networks in Table.~\ref{tab:comparisons}. A summary of mathematical notations is also given in Table.~\ref{tab:notations}.
This network consists of three crucial ingredients, i.e., the preprocessing layers, the capsule layers, and the quantum dynamic routing process.
The model's input is first fed into the preprocessing layers to extract some preliminary features. These features are encapsulated into several quantum states and then sent to the capsule layers.
Inside each capsule, there are a group of interacting qubits building up a sub-quantum neural network (sub-QNN). As such, given the $i$-th quantum capsule state $\chi^l_i$ in the $l$-th layer as the input, the parameterized sub-QNN can be regarded as a quantum channel $\mathcal{E}_{ij}$ that generates a prediction quantum state $\rho^{l+1}_{j|i} = \mathcal{E}_{ij} (\chi^l_i)$ for the $j$-th capsule in the $(l+1)$-th layer.
In order to obtain the $j$-th quantum capsule state in the $(l+1)$-th layer, $\chi^{l+1}_j = \sum_{i} q_{ij} \rho^{l+1}_{j|i}$, we propose a quantum dynamic routing algorithm. Here, $q_{ij}$ is the routing probability and dynamically updated with the geometric relationship (distance measure) between $\rho^{l+1}_{j|i}$ and $\chi^{l+1}_j$.  {As we will show below, such a quantum dynamic routing algorithm is able to evaluate the distance measure of quantum states.} The classification results can be read out by measuring the activation probability of capsules in the last layer.

There are several ways to measure the distance of quantum states \cite{Nielsen2010Quantum}, including the trace distance, fidelity, and the quantum Wasserstein distance \cite{Chakrabarti2019Quantum,Kiani2022Learning}. Here, we utilize the $k$-th moment overlap between the two mixed quantum states
\begin{equation}
	\label{eqn:fidelity}
	\Omega (\rho, \chi) \equiv \mathrm{Tr}(\rho^k \chi^k) \,,
\end{equation}
as a geometric measurement tool, where $k$ denotes the power of the matrix multiplication. The moment order $k$ serves as a hyperparameter in controlling the convergence of the iteration: the larger $k$ we choose, the quicker convergence we obtain.

During the quantum dynamic routing process, the overlap between the prediction state and the capsule state $\Omega (\rho^{l+1}_{j|i}, \chi^{l+1}_j)$ serves as a quantum geometric agreement.
Furthermore, the activation probability of each capsule can be accessed through the $k$-th purity of the state: 
\begin{equation} \label{eqn:purity}
    \mathcal{P}(\chi) \equiv \Omega (\chi, \chi) = \mathrm{Tr}(\chi^{2k}) \,.
\end{equation}
As the number of qubits increases, the minimum of the purity will exponentially decay to zero. 
In addition, since the prediction quantum state $\rho^{l+1}_{j|i} = \mathcal{E}_{ij} (\chi^l_i)$ heavily relies on the quantum process $\mathcal{E}_{ij}$ inside each capsule, the specific structure of sub-QNNs can have an enormous impact on the performance of QCapsNets. We will investigate several QCapsNets with different capsule architectures in the following paragraphs.

\subsection{Quantum dynamic routing}

We now introduce a quantum dynamic routing process among all the $M$ prediction states $\{ \rho^{l+1}_{j|i} \}_{i=1}^M$ and the $j$-th capsule state $\chi^{l+1}_j$ in $(l+1)$-th layer, which contains the following two major subroutines. 

\textit{(i) Measuring the geometric relationship in parallel.} In order to leverage the quantum parallelism, we first encode the set of prediction states $\{ \rho^{l+1}_{j|i} \}_{i=1}^M$ into the bus registers of two qRAM (quantum random access memory) states \cite{Giovannetti2008Quantum,Giovannetti2008Architectures,Park2019CircuitBased,Hann2019HardwareEfficient}, namely, the input qRAM state $\sigma_j^\mathrm{in} = 1/M \sum_{i=1}^M \ket{i} \bra{i} \otimes (\rho^{l+1}_{j|i})^{\otimes k}$ and the output qRAM state $\sigma_j^\mathrm{out} = \sum_{i=1}^M q_{ij} \ket{i} \bra{i} \otimes (\rho^{l+1}_{j|i})^{\otimes k}$.
The routing coefficients $q_{ij}$ of $\sigma_j^\mathrm{out}$ and $\chi^{l+1}_j$ are uniformly initialized at $1/M$ and then dynamically refined by geometric relationships. 
Owing to the qRAM structure, we can compute all the $k$-moment overlaps $\{ \Omega (\rho^{l+1}_{j|i}, \chi^{l+1}_j) \}_{i=1}^M$ in parallel by use of a SWAP test between $\sigma^\mathrm{in}_{j}$ and $\chi^{l+1}_j$ \cite{Buhrman2001Quantum}. Through tracing out the bus register, these geometric relationships can be efficiently encoded into the address register as a diagonal quantum density matrix $\omega_j = \sum_{i=1}^M \omega_{ij} \ket{i} \bra{i} $ with
\begin{equation} \label{eqn:RoutingCoefficients}
    \omega_{ij} = \frac{1+\Omega (\rho^{l+1}_{j|i}, \chi^{l+1}_j)}{\mathcal{A}} \, , \quad
    \mathcal{A} = \sum_{i=1}^M \left[ 1+\Omega (\rho^{l+1}_{j|i}, \chi^{l+1}_j) \right]
    \, ,
\end{equation}

\textit{(ii) Assigning the routing coefficients.} With such an overlap state $\omega_j$ at hand, we can utilize techniques of density matrix exponentiation \cite{Lloyd2014Quantum} and Hamiltonian simulation \cite{Berry2007Efficient} to generate the following unitary operator,
\begin{equation}
	U_j = \exp \left(-\mathrm{i} \frac{t}{\mathcal{A}} \sum_{i=1}^M \, \Omega (\rho^{l+1}_{j|i}, \chi^{l+1}_j) \, \ket{i} \bra{i} \otimes \mathbbm{1} \otimes X \right) \, ,
\end{equation}
where $\mathbbm{1}$ is the identity operator in the bus register, the Pauli-$X$ gate acts on the ancilla qubit, and $\mathcal{A}$ is a normalization factor defined in Eq.~\eqref{eqn:RoutingCoefficients}. Next, we apply $U_j$ to the state $\sigma_j^\mathrm{in} \otimes \ket{0} \bra{0}$ and project the above ancilla qubit to the $\ket{1} \bra{1}$ subspace via post-selection \cite{Lloyd2013Quantum}. As a consequence, all these $M$ geometric relationships can be assigned into the routing coefficients of the output qRAM state $\sigma_j^\mathrm{out}$ as
\begin{equation}
	q_{ij} \leftarrow \frac{|\Omega (\rho^{l+1}_{j|i}, \chi^{l+1}_j)|^2}{\sum_{i=1}^M|\Omega (\rho^{l+1}_{j|i}, \chi^{l+1}_j)|^2} \,.
\end{equation}
Subsequently, the new capsule state $\chi_j^{l+1}$ can be obtained by tracing out the index qubits of the output qRAM state $\mathrm{Tr_{index}} (\sigma_j^\mathrm{out}) = \sum_{i=1}^M q_{ij} (\rho^{l+1}_{j|i})^{\otimes k}$. Such a top-down feedback increases the routing coefficient $q_{ij}$ for the prediction state $\rho^{l+1}_{j|i}$ that has a large overlap with the new capsule state $\chi_j^{l+1}$. Repeating the above quantum routing procedure for a few iterations (usually three in our numerical experiments), the routing coefficients $q_{ij}$ generally converge.  {The explicit implementation and technical details of the algorithm are presented in the Appendix \ref{sec:Appx_Routing}.}

\section{Numerical experiments}
\begin{figure*}[t]
	\includegraphics[width=0.98\textwidth]{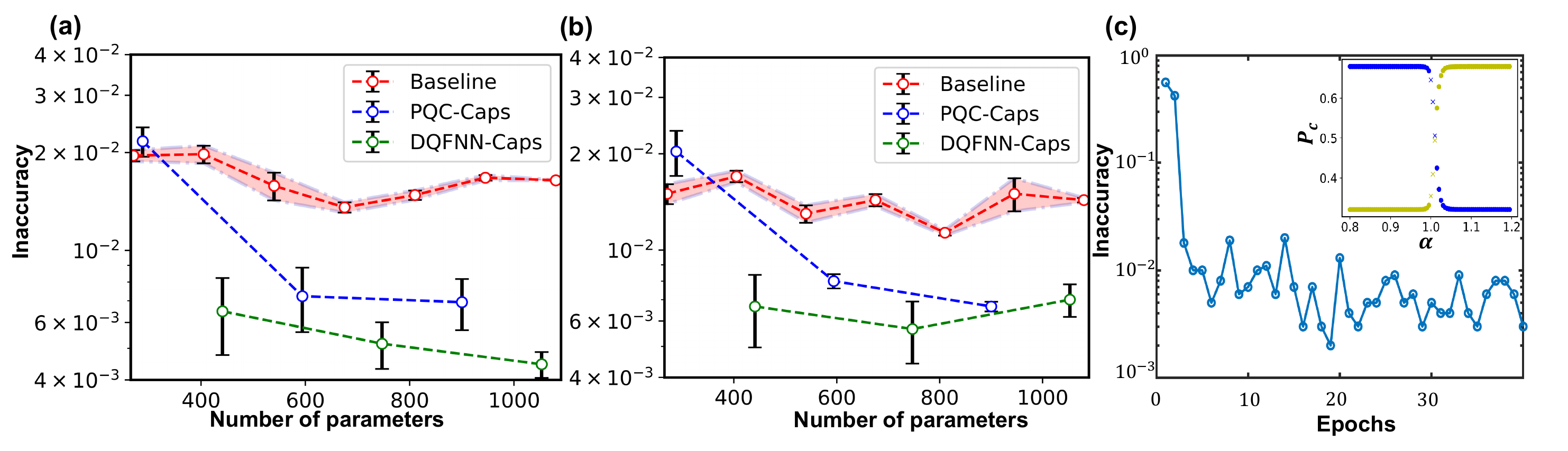}
    \caption{Numerical results of QCapsNets for classifications of handwritten-digit images and symmetry-protected topological states. (a)-(b) show the inaccuracy of the MNIST training and test datasets after the convergence, respectively, as a function of the number of parameters. For concrete comparisons, we equip the QCapsNets with two different sub-QNNs inside the capsule, namely the parameterized quantum circuit (PQC) [see the structure in Fig.~\ref{fig:structures}(a)], and the deep quantum feed-forward neural network (DQFNN) [where the structure is represented in Fig.~\ref{fig:structures}(b)]. Their corresponding QCapsNets are dubbed PQC-Caps and DQFNN-Caps, respectively. The baseline for comparison is a parameterized quantum circuit without any capsule architecture 
	[see the PQC structure in Fig.~\ref{fig:structures}(a)]. Each point refers to the averaged convergence results of three replicated trials, and the error bar indicates one standard deviation. (c) shows the training inaccuracy of QCapsNets for symmetry-protected topological states with respect to the training epochs.  Here we use the DQFNN-Caps structure. The training data contains $20000$ states, which are generated by uniformly sampling the model's parameter $\alpha$ from $0$ to $2$. We find that after $40$ epochs, the inaccuracy of the training dataset can be less than $10^{-2}$. The activation probabilities $P_c$ of two output capsules are plotted in the inset by varying $\alpha$ from 0.8 to 1.2. The $\circ$ dot indicates a right classification result while the $\times$ dot refers to a wrong one. This result shows that our trained QCapsNet can locate the quantum phase transition point near $\alpha=1$. }
	\label{fig:numerical_k}
\end{figure*}

\subsection{Performance benchmarking}
\label{performance_benchmarking}

To benchmark the performance of QCapsNets, we  carry out some numerical experiments about the classification of both classical (e.g., handwritten digit images) and quantum (e.g., topological states) data. 
Note that in QCapsNets, we can furnish their capsules with various sub-QNNs $\mathcal{E}_{ij}$.  Different families of sub-QNNs may bear distinct entangling capabilities and representation power. Thereby in the following numerical experiments, we propose two kinds of QCapsNets with different sub-QNNs, and then benchmark their performance by the classification accuracy. The first sub-QNN is the parameterized quantum circuit (PQC), which has been widely used as a standard ansatz for quantum classifiers \cite{Cerezo2021Variational,Benedetti2019Parameterized,Farhi2018Classification,Grant2018Hierarchical,Lu2020Quantum,Lu2021Markovian,Li2021Quantum}. The second one is the deep quantum feed-forward neural network (DQFNN), which has been proposed to solve the supervised learning problem \cite{Beer2020Training} and the dynamics of quantum open system \cite{Liu2022Solving}. Inside DQFNN, each node represents a qubit, and their connections are given by parameterized unitary gates.  {The explicit structures of these two sub-QNNs are shown in the Appendix~\ref{sec:Appx_Structure}.} For brevity, we name QCapsNets equipped with the above two sub-QNNs as PQC-Caps and DQFNN-Caps, respectively. In addition, we use the conventional parameterized quantum circuit (without any capsule architecture) as a baseline for comparison. We focus on the two-category classification problem, and thus supply the last layer of QCapsNets with two capsules,  {whose explicit structures are discussed in the Appendix~\ref{sec:Appx_Numerics}.}

We first apply QCapsNets to the classification of handwritten digit images in the MNIST dataset \cite{LeCun1998MNIST}, {which has been widely considered to be a   good ``toy model" for various machine learning paradigms.} In Fig.~\ref{fig:numerical_k}(a)-(b), we plot the scaling of inaccuracy as a function of the number of parameters in QCapsNets for the training and test dataset, respectively. {  As the number of parameters increases, the averaged inaccuracy of both QCapsNets is reduced to less than $2 \times 10^{-2}$, which surpasses the performance of the baseline. } Therefore, given the same number of parameters, owing to the capsule architecture and the quantum dynamic routing mechanism, QCapsNets can extract the information more effectively and may possess an enhanced representation power than the conventional parameterized quantum circuits { [see the specific structure of the parameterized quantum circuit in Fig.~\ref{fig:structures}(a)]}. 

\begin{figure*}[t]
	\includegraphics[width=0.98\textwidth]{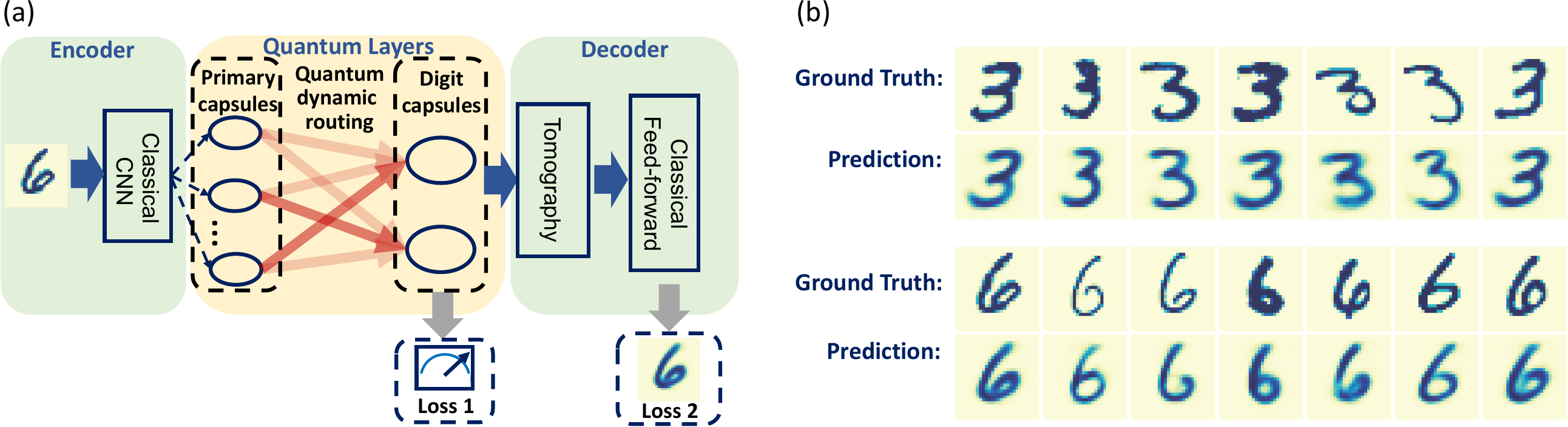}
 	\caption{ Reconstruction of handwritten-digit images with QCapsNets.  (a) The architecture of the reconstruction networks. We use a classical CNN to extract some preliminary features of the input image, and then encode its output into nine primary capsules in the quantum layers. Through quantum dynamic routing process, the essential features of input images are encoded into two digit capsules.  {Each capsule in the quantum layers is composed of four qubits. We select the capsule with the largest activation probability [Eq.~\eqref{eqn:purity}] for quantum state tomography, and feed its classical representation to the classical feed-forward network to reconstruct the input image. }The loss function of this model includes two parts. The first part requires the measurement results of the digit capsules, and the second part evaluates the reconstruction error of the decoder networks. (b) The reconstruction results with QCapsNets. Here, we present images with two different labels, i.e., ``3'' and ``6''. For comparison, we show the input (ground truth) data in the first row and the reconstruction result (prediction) in the second row.}
	\label{fig:reconstruction}
\end{figure*}

In addition to classical data, QCapsNets also apply to quantum input data. To this end, we further use the QCapsNet as a quantum classifier for the symmetry-protected topological (SPT) states. Specifically, we consider the following cluster-Ising model, whose Hamiltonian reads \cite{Smacchia2011Statistical}:
\begin{equation}
    H(\alpha) = -\sum_{j=1}^{n} X_{j-1} Z_{j} X_{j+1} + \alpha \sum_{j=1}^{n} Y_j Y_{j+1},
\end{equation}
where the Pauli matrices $X_j,Y_j,Z_j$ act on the $j$-th spin and $n$ is the total number of spins. The parameter $\alpha$ indicates the strength of the nearest coupling. This model is exactly solvable and exhibits a well-understood quantum phase transition point at $\alpha=1$. There is an antiferromagnetic phase for $\alpha>1$, while a $\mathbb{Z}_2 \times \mathbb{Z}_2$ SPT phase for $\alpha<1$ (characterized by a non-local string order). The training data is a set of ground states of Hamiltonian $H(\alpha)$, which is generated by uniformly sampling $\alpha$ from $0$ to $2$ under the periodic boundary condition. In this example, the capsule structure of our QCapsNet is fixed to be DQFNN. 

As shown in Fig.~\ref{fig:numerical_k}(c), the inaccuracy of the training dataset can drop below $10^{-2}$ within 40 epochs. After training, we generate $80$ ground states as the test dataset, with $\alpha$ ranging from $0.8$ to $1.2$. In the inset, we feed the trained QCapsNets with the test dataset, and plot the activation probability $P_c$ of two output capsules as a function of $\alpha$. In the regime far from the critical point, as the magnitudes of $P_c$ significantly depart away, our QCapsNet can precisely distinguish two different phases. In addition, the phase transition point can be inferred from the intersection of $P_c$. Although there are some tiny fluctuations near the quantum phase transition point due to finite-size effects, the critical point estimated by the QCapsNet is $1.01$, i.e., only a small deviation from the exact value.  {The loss function and the explicit activation probability of capsules are given in the Appendix~\ref{sec:Appx_Numerics}.}

\subsection{Visualize the capsule subspace}
\begin{figure*}[t]
	\includegraphics[width=0.98\textwidth]{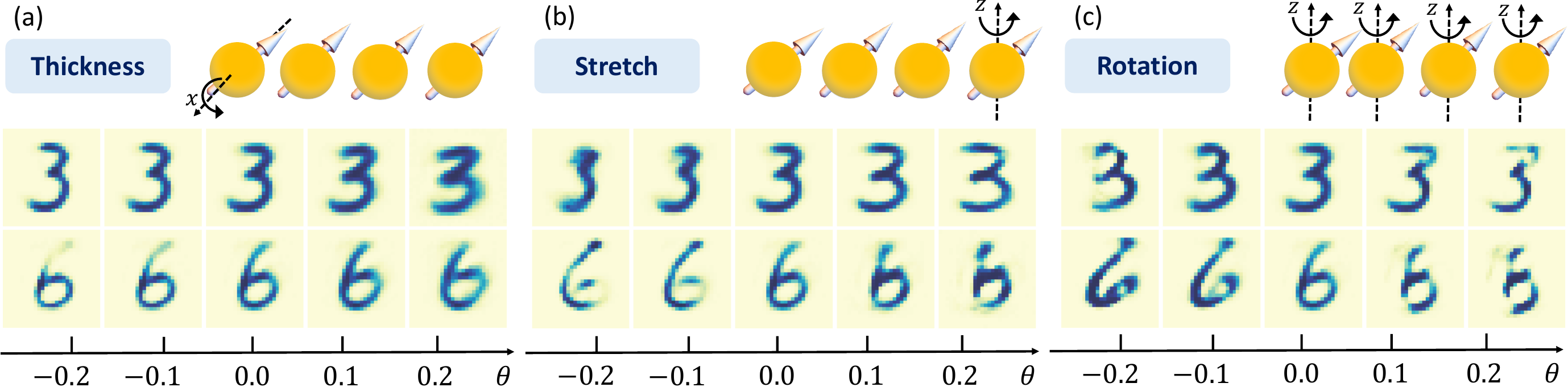}
	\caption{ Variations of handwritten-digit images with different perturbations on the active digit capsule. (a) We apply a $x$-axis rotation $U = \mathrm{e}^{-\mathrm{i} \theta X_1 }$ on the first qubit in the active digit capsule, which shows a variation of the thickness in the reconstruction images. (b) We apply a $z$-axis rotation $U = \mathrm{e}^{-\mathrm{i} \theta Z_4}$ on the last qubit in the active digit capsule, which indicates a stretch behavior in the reconstruction images. (c) We apply a global $z$-axis rotation $U = \mathrm{e}^{-\mathrm{i} \theta Z_1 Z_2 Z_3 Z_4}$ on the active digit capsule. The reconstruction images are rotated in different angles, along with a slight deformation. The perturbation parameter $\theta$ is tweaked from $-0.2$ to $0.2$ by intervals of 0.06.}
	\label{fig:interpretion_perturbation}
\end{figure*}

 {We have demonstrated the enhanced performance of QCapsNets in the classification tasks.} Yet, if one would like to further utilize QCapsNets for making critical decisions (e.g., self-driving cars and medical diagnostics) \cite{Finlayson2019Adversarial}, it is of crucial importance to understand, trust, and explain the rational behind the model's decisions.  {In the field of computer vision and machine learning, explainability aims to investigate how the parameters of neural networks affect the classification results in a human-understandable way. For instance, the heatmap (class activation mapping) enables us to visualize which feature dominates the final prediction in CNNs \cite{Selvaraju2020GradCAM}. Likewise, we would like to explore the potential explainability of QCapsNets through the following reconstruction scheme. We remark that our goal is to investigate whether QCapsNets could extract some explainable information from the input data. Hence, we adopt a classical-quantum hybrid architecture to alleviate the limitation of classically simulating quantum neural networks. } 

In Fig.~\ref{fig:reconstruction}(a), we attach the QCapsNet to a classical encoder (CNN) and a classical decoder (feed-forward network), and use the whole network to reconstruct the input image from the MNIST dataset.  The first two procedures are similar to the ones in the classification task. Some basic features of the image are first extracted by the classical encoder, and then encapsulated into nine primary capsules in the quantum layers. Through quantum dynamic routing process, high-level features are encoded into two digit capsules. We pick up the capsule state with the largest activation probability, and feed it into the classical feedforward network to reconstruct the input image. {  In order to obtain a classical representation of the chosen capsule state $\chi$, we perform quantum state tomography by measuring the observables under the Pauli basis, i.e. $\mathrm{Tr}(\chi \sigma_{x,y,z,I})$, respectively. }
To guide the active capsule to capture more intrinsic features, we use a composite loss function which takes into account both the classification and reconstruction loss  {(see the Appendix~\ref{sec:Appx_Numerics} for their explicit expressions).} After training, the reconstruction results are plotted in Fig.~\ref{fig:reconstruction}(b), where the first row shows the input images (ground truth) and the second row exhibits the reconstruction images (prediction). These reconstruction results are considerably robust while preserving their representative details. We note in passing that such a classical-quantum hybrid architecture shown in Fig.~\ref{fig:reconstruction}(a) is just for a proof-of-principle demonstration of the potential explainability. In the future, we could replace the classical neural network with a quantum generative model, and design an end-to-end quantum reconstruction model to reduce the cost of quantum state tomography.

In the above simulation, we find that the most active capsule contains sufficient information to reconstruct the original image. Hence, the entire Hilbert space of such a quantum capsule state may learn a plenty of variants of the images. As shown in Refs.~\cite{Sabour2017Dynamic,Shahroudnejad2018Improved}, adding perturbations into the capsule is one of useful tools to reveal the potential explainability for each subspace. After the reconstruction process, we can feed a perturbed capsule state to the trained decoder network, and analyze how the perturbation affects the reconstruction result. As shown in Fig.~\ref{fig:interpretion_perturbation}, we test three different types of perturbations on the digit capsule, with the perturbation parameter $\theta$ ranging from $-0.2$ to $0.2$. These perturbations are applied on the digit capsules before we measure them under the Pauli basis. The first type of perturbation we consider is the $x$-axis rotation $U =  \mathrm{e}^{-\mathrm{i} \theta X_1 }$ on the first qubit. In Fig.~\ref{fig:interpretion_perturbation}(a), as the perturbation parameter $\theta$ gets larger, the strokes of both digits ``3'' and ``6'' become thicker. In Fig.~\ref{fig:interpretion_perturbation}(b), we apply a $z$-axis rotation $U = \mathrm{e}^{-\mathrm{i} \theta Z_4}$ on the last qubit. Through tuning the perturbation parameter, both digits have been squeezed to various degrees. In the Fig.~\ref{fig:interpretion_perturbation}(c), we apply a global $z$-axis rotation $U = \mathrm{e}^{-\mathrm{i} \theta Z_1 Z_2 Z_3 Z_4}$ on the whole active capsule. By tweaking the perturbation parameter, both digits are rotated at different angles, together with a tiny deformation. These perturbation results indicate that a particular subspace of the digit capsule state could almost represent a specific explainable feature of the handwritten images.

\section{Conclusion}

 {We have introduced a QCapsNet model equipped with an efficient quantum dynamic routing algorithm, where the overlaps between two capsule states can be obtained in parallel. Through the classification tasks for both the classical handwritten digits and symmetry-protected topological states, we found that QCapsNets achieve an enhanced accuracy among quantum classifiers we considered.} By virtue of the geometric relationships, QCapsNets can capture the essential features of the input data, and then encapsulate them into the output capsule. Accordingly, such capsule contains sufficient instantiation parameters to reconstruct the original data. In particular, one specific subspace of the output capsule could correspond to a human-understandable feature of the input data. %

Many interesting and important questions remain unexplored and deserve further investigation.  First, it would be great interesting to consider different experimental scheme to quantify the geometric relationships between capsules states. Indeed, there are several works show that such a quantity can be measured with cross-platform  verification \cite{Elben2020CrossPlatform,Anshu2022Distributed}, which do not require a quantum communication and much suitable for the noisy intermediate-scale quantum (NISQ) devices \cite{preskill2018quantum}. Second, a quantum neural networks may suffer by the barren plateaus problem \cite{Mcclean2018Barren,Wang2021Noiseinduced,Cerezo2021Higher,Sharma2022Trainability,Arrasmith2021Effect,Holmes2021Barren,OrtizMarrero2021EntanglementInduced,Patti2021Entanglement,Pesah2021Absence,Arrasmith2022Equivalence,Cerezo2021Cost,Uvarov2021Barren,Grant2019Initialization,Zhao2021Analyzing,liu2021presence}, where the gradient of parameters may vanish in any direction. A good choice of the initial guess of the parameters should be important for the maintain the trainability of QCapsNets. Third, the scalability and performance of QCapsNets may be further improved by other quantum routing algorithms. For instance, a quantum extension of the expectation-maximization routing algorithm \cite{Hinton2018Matrix, Miyahara2020quantum, kerenidis2020quantum} might be a good candidate along this line.
In addition, quantum adversarial machine learning has attracted considerable attention recently \cite{Lu2020Quantum,Liu2020Vulnerability}.  It has been shown that, compared to CNNs, CapsNets are relatively resilient against the adversarial attacks \cite{Hinton2018Matrix}. This inspires a natural question concerning whether QCapsNets are also more robust to adversarial perturbations than other traditional quantum classifiers.  {  Furthermore, we may exploit quantum neural tangent kernels to understand the training dynamics of QCapsNets \cite{Liu2022Representation}. Recently, we notice a remarkable paper proposing a dequantized version of the SWAP test \cite{tang2021quantum}. 
We may try to design some efficient quantum-inspired algorithms for CapsNets.}
Finally, such a grouped architecture may shed light on implementing QCapsNets in a collection of distributed quantum computers that are routed via the quantum internet \cite{Kimble2008Quantum}.

\section{Acknowledgements}

We thank Sirui Lu, Jin-Guo Liu, Junyu Liu, and Haoran Liao for helpful discussion. This work was supported by the Frontier Science Center for Quantum Information of the Ministry of Education of China, Tsinghua University Initiative Scientific Research Program, and the Beijing Academy of Quantum Information Sciences. D.-L. D. also acknowledges additional support from the Shanghai Qi Zhi Institute. Certain images in this publication have been obtained by the author(s) from the Wikimedia website, where they were made available under a Creative Commons license or stated to be in the public domain.  Please see individual figure captions in this publication for details. To the extent that the law allows, IOP Publishing disclaim any liability that any person may suffer as a result of accessing, using or forwarding the image(s).  Any reuse rights should be checked and permission should be sought if necessary from Wikipedia/Wikimedia and/or the copyright owner (as appropriate) before using or forwarding the image(s).

\appendix

\section{Classical capsule networks}

\label{sec:Appx_CapsNets}

\begin{figure}[b]
    \centering
    \includegraphics[width=1\textwidth]{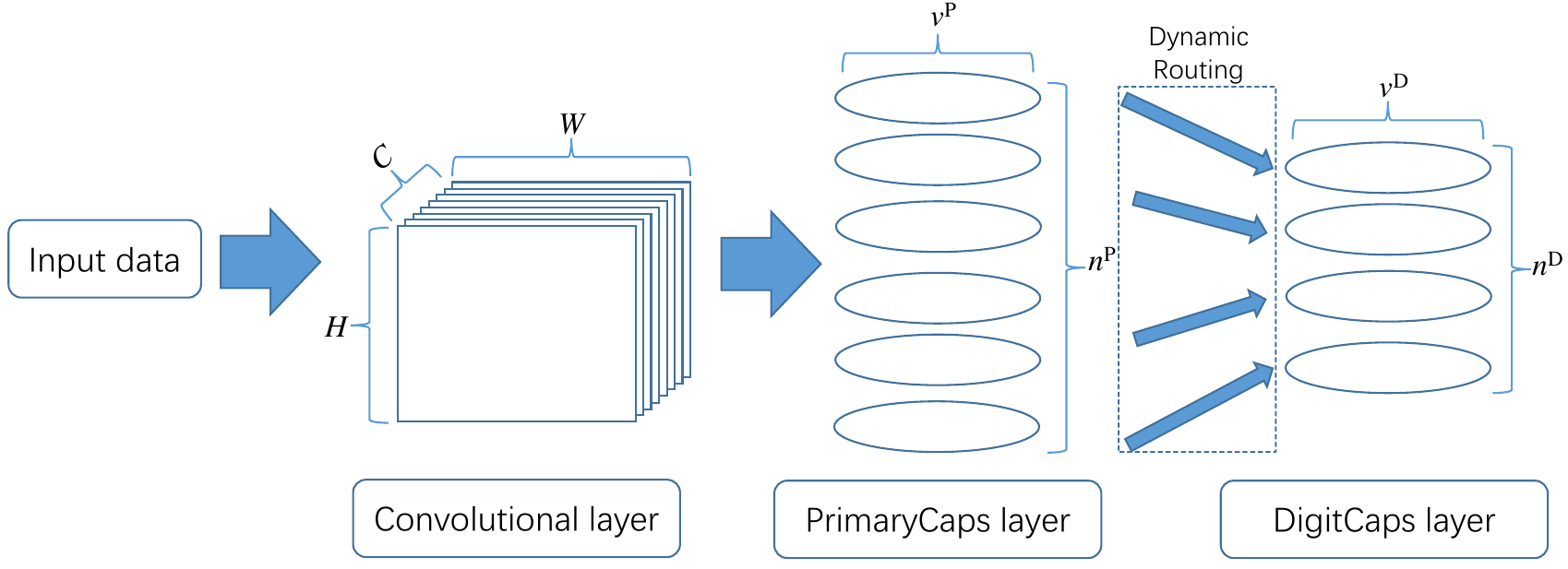}
    \caption{The overall structure of classical capsule networks with three essential layers. The input data are first processed through the convolutional layer. Then the corresponding feature maps $[H,W,C]$ are fed into the PrimaryCaps layer and reshaped into $n^\textsc{p}$ capsules living in a $v^\textsc{p}$-dimensional Euclidean space. Owing to the dynamic routing algorithm, the DigitCaps layer ($n^\textsc{d}$ capsules in a $v^\textsc{d}$-dimensional space) is able to extract features of entities from lower-level capsules and capture their geometric relationships. The norm of the active capsule in the DigitCaps layer indicates the probability of a specific class being detected.}
    \label{fig:capnet2}
\end{figure}

\subsection{General architecture} 

The classical capsule network (CapsNet) was first introduced in Ref.~\cite{Hinton2011Transforming}, and then equipped with a dynamic routing algorithm in Ref.~\cite{Sabour2017Dynamic} and an expectation-maximization (EM) routing algorithm in Ref.~\cite{Hinton2018Matrix}. CapsNets are multi-layered networks, whose building block is a capsule that represented by a \textit{vector}, instead of a \textit{scalar} in Vanilla neural networks (see Table.~\ref{tab:comparisons} for their detailed comparisons) \cite{Liao2021CapsNetTensorflow,Geron2017HandsOn}. The norm (length) of the vector reflects the probability of an \textit{entity} being present, while the components (orientations) of the vector indicate the \textit{features} of the entity. In computer graphics \cite{Hughes2013Computer}, features refer to different types of instantiation parameters of an entity, such as its rotation, pose, hue, texture, deformation, etc. The capsule can learn to detect an entity in an object, together with its corresponding features.

The structure of CapsNets generally consist of three parts (as shown in Fig.~\ref{fig:capnet2}). The first part is the convolutional layer that used in convolutional neural networks (CNNs). In this layer, the model detects the basic local features of the input data. For the specific task demonstrated in Ref.~\cite{Sabour2017Dynamic}, the MNIST handwritten data are processed through two convolutional layers: $[28,28,1] \xrightarrow[\text{kernel}: 9\times 9 \text{, stride}: 1]{\text{ReLU Conv2d}} [20,20,256] \xrightarrow[\text{kernel}: 9\times 9 \text{, stride}: 2]{\text{ReLU Conv2d}} [6,6,256] = [6,6,32 \times 8]$. The second part is the PrimaryCaps layer, which not only captures the features of data, but also generates their combinations. In this layer, the above feature map is reshaped into $[n^\textsc{p}, v^\textsc{p}] = [6 \times 6 \times 32,8]$, corresponding to 1152 capsule vectors living in the $8$-dimensional Euclidean space. These capsule vectors are then fed into the third part --- the DigitCaps layer. Therein, via dynamic routing between the PrimaryCaps layer and the DigitCaps layer, the shape of the final output vector is $[n^\textsc{d}, v^\textsc{d}] = [10, 16]$, where $n^\textsc{d}$ is equal to the number of categories for classification. The probability can be read from the norm of the vector, while its components represent the features of the entity, which can be further used to unbox the learning model and seek for its explainability.

\begin{algorithm}[t]
    \label{alg:ClassicalDynamicRouting}
    \SetKwInOut{Input}{input}\SetKwInOut{Output}{output}

    \Input{the index of layer $l$, the number of iteration $e$, the prediction vector $\mathbf{u}^{l+1}_{j|i} = \mathbf{W}_{ij} \mathbf{v}^l_{i}$}
    \Output{the capsule vector $\mathbf{v}^{l+1}_j$ in the $(l+1)$-th layer }
    \textbf{initialize}: $\forall i \in \mathcal{L}_{l}, j \in \mathcal{L}_{l+1}: b_{ij} \leftarrow 0$ \tcp*[f]{initialize a uniform routing path}\\
    \For{$e$ iterations}{
        $\forall i \in \mathcal{L}_{l}, j \in \mathcal{L}_{l+1}: r_{ij} \leftarrow \operatorname{softmax}_{j} \left ( b_{ij} \right ) $ \tcp*[f]{$\mathrm{softmax}_j$ computes Eq.\eqref{eqn:softmax}}\\
        $\forall j \in \mathcal{L}_{l+1}: \mathbf{s}^{l+1}_j \leftarrow \sum_{i} r_{ij} \mathbf{u}^{l+1}_{j|i}$ \tcp*[f]{weight with the routing parameters}\\
        $\forall j \in \mathcal{L}_{l+1}: \mathbf{v}^{l+1}_{j} \leftarrow \operatorname{squash} \left (\mathbf{s}^{l+1}_{j} \right )$ \tcp*[f]{$\mathrm{squash}$ computes Eq.\eqref{eqn:squash}}\\
        $\forall i \in \mathcal{L}_{l}, j \in \mathcal{L}_{l+1}: b_{ij} \leftarrow b_{ij} + \mathbf{u}_{j|i}^{l+1} \cdot \mathbf{v}_j^{l+1}.$ \tcp*[f]{update with the geometric relation}
    }
    \Return{$\mathbf{v}^{l+1}_j$}
    \caption{Dynamic routing for classical capsule networks}
\end{algorithm} 

\begin{figure}[t]
    \centering
    \includegraphics[width=0.70\textwidth]{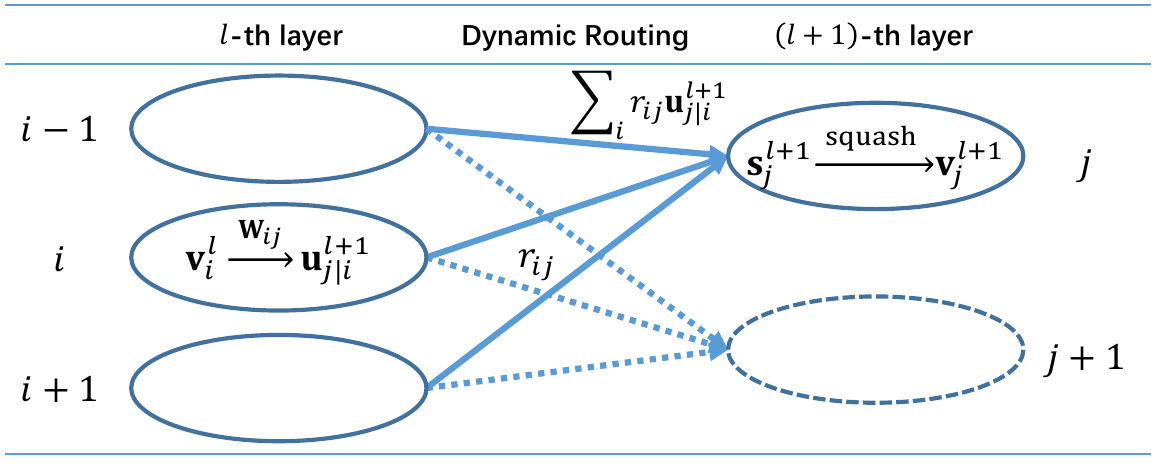}
    \caption{Illustration of the classical dynamic routing. The prediction vector $\mathbf{u}^{l+1}_{j|i}$ of the $j$-th capsule in the $(l+1)$-th layer is calculated by $\mathbf{u}^{l+1}_{j|i} = \mathbf{W}_{ij} \mathbf{v}^l_{i}$, where $\mathbf{W}_{ij}$ is a weight matrix and $\mathbf{v}^l_{i}$ is the $i$-th capsule in the $l$-th layer. The unnormalized capsule $\mathbf{s}_j^{l+1}$ is given by $\mathbf{s}_j^{l+1} = \sum\nolimits_{i} r_{ij} \mathbf{u}_{j|i}^{l+1}$ with routing coefficients $r_{ij}$. The output capsule is obtained by $\mathbf{v}^{l+1}_j = \mathtt{squash}(\mathbf{s}_j^{l+1})$ with a nonlinear function $\texttt{squash}$.}
    \label{fig:capnet}
\end{figure}

\subsection{Dynamic routing} 

In the following context we will elucidate the dynamic routing algorithm \cite{Sabour2017Dynamic}, whose pseudocode is presented in Algorithm~\ref{alg:ClassicalDynamicRouting}. Consider the dynamic routing algorithm between the $l$-th and the $(l+1)$-th capsule layer in the classical CapsNets (as shown in Fig.~\ref{fig:capnet}). Given the $i$-th capsule $\mathbf{v}^l_i$ in the $l$-th layer, we first calculate the prediction vector $\mathbf{u}^{l+1}_{j|i}$ of the $j$-th capsule in the $(l+1)$-th layer by multiplying a weight matrix $\mathbf{W}_{ij} \in \mathbb{R}^{v_\textsc{d} \times v_\textsc{p}}$ (trained by backpropagation),
\begin{equation}
    \mathbf{u}^{l+1}_{j|i} = \mathbf{W}_{ij} \mathbf{v}^l_{i},  \quad \forall i \in \mathcal{L}_l, j \in \mathcal{L}_{l+1} \,.
\end{equation}
Then we sum over all the prediction vectors $\mathbf{u}^{l+1}_{j|i}$ in the $l$-th layer with routing coefficients $r_{ij}$ to generate the pre-squashed vector $\mathbf{s}_j^{l+1}$ of the $j$-th capsule in the $(l+1)$-th layer,
\begin{equation}
    \mathbf{s}_j^{l+1} = \sum\nolimits_{i} r_{ij} \mathbf{u}_{j|i}^{l+1} \,,
\end{equation}
where $r_{ij}$ quantifies the probability that the capsule vector $i$ may affect the capsule vector $j$. Finding groups of similar $\mathbf{u}_{j|i}^{l+1}$ can be seen as a clustering problem, thus we can interpret $r_{ij}$ as the probability that capsule $i$ is grouped to cluster $j$, which requires $\sum_j r_{ij} = 1$. To this end, we introduce a set of digits (unnormalized routing coefficients) $b_{ij}$, which is related to the routing coefficient $r_{ij}$ by a softmax function:
\begin{equation} \label{eqn:softmax}
    r_{ij} = \mathtt{softmax}_j(b_{ij}) \equiv \frac{\exp(b_{ij})}{\sum_j\exp(b_{ij})} \,.
\end{equation}
We remark that the nonlinearity of the classical CapsNets is contributed by the so-called ``squashing'' operation:
\begin{equation} \label{eqn:squash}
    \mathbf{v}_j^{l+1} = \mathtt{squash}(\mathbf{s}_j^{l+1}) \equiv \frac{\|\mathbf{s}_j^{l+1}\|^2}{1+\|\mathbf{s}_j^{l+1}\|^2} \frac{\mathbf{s}_j^{l+1}}{\|\mathbf{s}_j^{l+1}\|}.
\end{equation}
The $\mathtt{squash}$ function forces short vectors to get shrunk to almost zero length, and long vectors to get shrunk to a length slightly below one, which is analogous to $\mathtt{sigmoid}(x) \equiv 1 / (1 + e^{-x})$ in the classical neural networks.
The essence of the dynamic routing algorithm is the so-called routing-by-agreement mechanism, which is able to exact the geometric relationship $g_{ij}$ between prediction vectors $\mathbf{u}^{l+1}_{j|i}$ and the possible output vector $\mathbf{v}^{l+1}_j$. Note that capsules vectors are living in the Euclidean space, the geometric relationship can be characterized by the dot product between them:
\begin{equation}
    g_{ij} = \mathbf{u}_{j|i}^{l+1} \cdot \mathbf{v}_j^{l+1}.
\end{equation}
At the beginning of the routing algorithm, all the routing digits $b_{ij}$ are initialized to zero, which leads to a uniformly routing path between layers. During iterations, the routing digits are then iteratively refined by the agreement $g_{ij}$ between the current output capsule $\mathbf{v}_j^{l+1}$ and the prediction vector $ \mathbf{u}_{j|i}^{l+1}$,
\begin{equation}
    b_{ij} \leftarrow b_{ij} + g_{ij}.
\end{equation}
Such a routing-by-agreement mechanism has been demonstrated far more effective than the max-pooling method used in CNNs \cite{Sabour2017Dynamic,Hinton2018Matrix}. In short, CapsNet is a neural network which replaces scalar-output neurons with vector-output capsules and max-pooling with routing-by-agreement.

\subsection{Training and reconstruction}

In order to detect possibly multiple handwritten digits in an image, the classical CapsNets use a separate margin loss $L_c$ for each class $c$ digit present in the image \cite{Sabour2017Dynamic}:
\begin{equation}
    L_c = T_c \max (0, m^+ - \| \mathbf{v}_c^\textsc{d} \|)^2 + \lambda (1-T_c) \max(0, \| \mathbf{v}_c^\textsc{d} \| - m^-)^2, 
\end{equation}
where $\mathbf{v}_c^\textsc{D}$ is the $c$-th capsule in the DigitCaps layer, $\text{max}(A,B)$ is the maximum value between $A$ and $B$, and $T_c$ is a indicator function for a specific class $c$, i.e., $T_c = 1$ if and only if an object of class $c$ is present. The margins are set as $m^+=0.9$ and $m^-=0.1$. Here, $\lambda$ is a down-weighting term that prevents the initial learning from shrinking the activity vectors of all classes. The total margin loss $L_\mathrm{M} = \sum_c L_c$ is just the sum of the losses of all classes. With such a loss function at hand, one can routinely utilize the backpropagation algorithm to train the parameters in the weight matrices $\mathbf{W}_{ij}$. We remark that the routing coefficients $r_{ij}$ are not trainable parameters, but values determined by the dynamic routing algorithm. 

Apart from the aforementioned three layers for classification, one can add additional reconstruction layers to enable capsules to encode the corresponding features of input data. Specifically, the most active capsule in the DigitCaps layer can be fed into a decoder which consist of three fully connected layers. After minimizing the mean square error $L_\mathrm{MSE}$ between the outputs of the logistic units and the original pixel intensities, one can use initiation parameters in the most active capsule to reconstruct the input image. In this vein, both the margin loss and the MSE loss should be optimized simultaneously, $L = L_\mathrm{M} + \gamma L_\mathrm{MSE}$, where $\gamma$ is a small regularization term that scales down $L_\mathrm{MSE}$ so that it does not dominate $L_\mathrm{M}$ during training.

\section{Implementation of the quantum dynamic routing algorithm}

\label{sec:Appx_Routing}

\begin{table*}[t]
    \setlength\extrarowheight{3pt}
        \caption{\label{tab:comparisons} Comparisons of different networks}
        \begin{tabularx}{\textwidth}{c|c|c|c|X}
            \cline{1-5}
            \multicolumn{2}{c|}{Networks}                          & Classical neural networks & Classical capsule networks & \makecell[c]{Quantum capsule networks} \\ \cline{1-5}
            \multicolumn{2}{c|}{Input}                             & scalar $x^l_i$ & vector $\mathbf{v}^l_i$ & \makecell[c]{quantum density matrix $\chi^l_i$ }\\ \cline{1-5}
            \multirow{3}{*}{Operations} & Transformation & $y^{l+1}_{j|i} = w_{ij} x^l_i + b^l_i$ & $\mathbf{u}^{l+1}_{j|i} = \mathbf{W}_{ij} \mathbf{v}^l_{i}$ & \makecell[c]{quantum sub-neural networks: \\ $\rho^{l+1}_{j|i} = \mathcal{E}_{ij} (\chi^l_i)$} \\ \cline{2-5} 
                                        & Weighting & $z^{l+1}_j = \sum\nolimits_i 1 \cdot y^{l+1}_{j|i}$ &  $\mathbf{s}^{l+1}_j  = \sum\nolimits_i r_{ij} \cdot \mathbf{u}^{l+1}_{j|i}$ & \makecell[c]{$\sigma^\mathrm{out}_j = \sum\nolimits_i q_{ij} \ketbra{i} \otimes (\rho^{l+1}_{j|i})^{\otimes k}$ }\\ \cline{2-5} 
                                        & Normalization & $x^{l+1}_j = \mathtt{sigmoid}(z^{l+1}_j)$ & $\mathbf{v}^{l+1}_j = \mathtt{squash}(\mathbf{s}^{l+1}_j)$ & \makecell[c]{trace out the index register: \\$(\chi^{l+1}_j)^{\otimes k} = \mathrm{Tr}_\mathrm{index}(\sigma^\mathrm{out}_j)$
                                        } \\ \cline{1-5}
            \multicolumn{2}{c|}{Output}                            & scalar $x^{l+1}_j$         & vector $\mathbf{v}^{l+1}_j$ & \makecell[c]{quantum density matrix $\chi^{l+1}_j$} \\ \cline{1-5}
        \end{tabularx}
\end{table*}

In this section, we will elucidate the quantum dynamic routing process among all the $M$ prediction states $\{ \rho^{l+1}_{j|i} \}_{i=1}^M$ and the $j$-th capsule state $\chi^{l+1}_j$ in $(l+1)$-th layer, whose pseudocode is presented in Algorithm~\ref{alg:QuantumDynamicRouting}. A concise comparison to the classical CapsNets is given in Table.~\ref{tab:comparisons}. For brevity, the $(l+1)$ superscript of the density matrix are omitted temporally in the following context.
The overall quantum circuit implementation for our quantum dynamic routing algorithm is demonstrated in Fig.~\ref{fig:sp_circuit}.

\subsection{Measure the geometric relationship in parallel}

Recall that in the main text, we quantify the geometric relationship between the prediction state $\chi_j$ and the capsule state $\rho_{j|i}$ as the $k$-th moment overlap:
\begin{equation}
    \Omega (\rho_{j|i}, \chi_j) = \mathrm{Tr}(\rho_{j|i}^k \chi_j^k) \,,
\end{equation}
where $k$ denotes the power of the matrix multiplication. To obtain all the $M$ overlaps $\{ \Omega (\rho_{j|i}, \chi_j) \}_{i=1}^M$ in parallel, we introduce a qRAM oracle $O_v$ to generate two qRAM states, $\sigma_j^\mathrm{in}$ and $\sigma_j^\mathrm{out}$, which correlate all the $M$ prediction states $\{ \rho_{j|i} \}_{i=1}^M$ as
\begin{align}
    \sigma_j^\mathrm{in} &= O_v \left[ \frac{1}{M} \sum_{i=1}^M |i\rangle \langle i| \otimes (|\vec{0}\rangle \langle \vec{0}|)^{\otimes k} \right] O_v^\dagger = \frac{1}{M} \sum_{i=1}^M |i\rangle \langle i| \otimes \rho^{\otimes k}_{j|i} \,, \label{eqn:qRAMin} \\
    \sigma_j^\mathrm{out} &= O_v \left[ \sum_{i=1}^M q_{ij} |i\rangle \langle i| \otimes (|\vec{0}\rangle \langle \vec{0}|)^{\otimes k} \right] O_v^\dagger = \sum_{i=1}^M q_{ij} |i\rangle \langle i| \otimes \rho^{\otimes k}_{j|i} \,, \label{eqn:qRAMout}
\end{align}
where $|i\rangle$ refers to the index (address) state, $\rho_{j|i}^{\otimes k}$ denotes $k$ copies of the prediction states stored in the bus register, and $q_{ij}$ is the routing probability from the prediction state $\rho_{j|i}$ to the capsule state $\chi_j$. Note that the routing probability of the input qRAM state $\sigma_j^\mathrm{in}$ is uniformly fixed at $1/M$ during the whole quantum routing process. However, the routing probability $q_{ij}$ of the output qRAM state $\sigma_j^\mathrm{out}$ is uniformly initialized to $1/M$ only in the first iteration, and then dynamically refined by the geometric relationship. 

\begin{figure}
\centering
\includegraphics[width=0.98\textwidth]{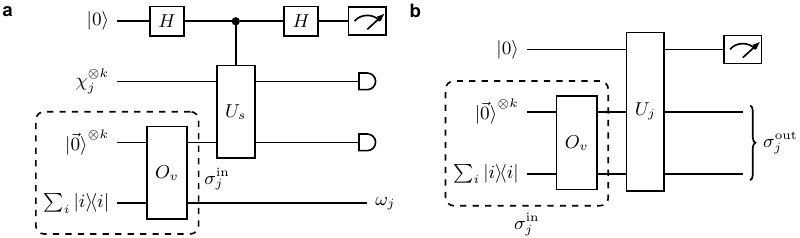}
\caption{Quantum circuits for the quantum dynamic routing algorithm. (a) Subroutine for parallel measuring the geometric relationship. The quantum circuit contains four registers: an index register initialized to $\sum_{i=1}^M \ketbra{i}$, a bus register initialized to $k$ copies of $|\vec{0}\rangle$, a capsule register initialized to $k$ copies of the capsule state $\chi_j$, and an ancilla register initialized to $|0\rangle$. The qRAM oracle $O_v$ will generate an input qRAM state $\sigma_j^\mathrm{in} = 1/M \sum_{i=1}^M |i \rangle \langle i | \otimes \rho^{\otimes k}_{j|i}$ that correlates $M$ prediction states $\{ \rho_{j|i} \}_{i=1}^M$ in the bus register. Then we implement a SWAP test by applying a controlled $2k$-SWAP gate [controlled-$U_s$ in Eq.~\eqref{eqn:2kSWAP}] to the ancilla, capsule and bus registers. In the output of the circuit, we measure the ancilla qubit and project it in the $|0\rangle$ subspace with a post-selection scheme. Finally, we trace out both the capsule and bus registers, and thus obtain the overlap density matrix $\omega_j = \sum_{i=1}^M \omega_{ij} |i\rangle \langle i |$, where $\omega_{ij} \propto 1+\mathrm{Tr}(\rho_{j|i}^k \chi_j^k)$ contains the geometric information. (b) Subroutine for assigning the routing probability to each capsule. Here we introduce another ancilla qubit that initialized to $|0\rangle$. The unitary operator is  {$U_j = \exp(-\mathrm{i} t / \mathcal{A} \sum_{i=1}^M \, \mathrm{Tr}(\rho_{j|i}^k \chi_j^k) \, |i\rangle \langle i| \otimes \mathbbm{1} \otimes X)$, where $\mathcal{A}$ is a normalization factor, $\mathbbm{1}$ is the identity operator in the bus register and the Pauli-$X$ gate acts on the ancilla qubit.} After a short evolution, we project the ancilla qubit in the $|1\rangle$ subsystem with a post selection scheme. The geometric relationships $q_{ij} \propto |\mathrm{Tr}(\rho_{j|i}^k \chi_j^k)|^2$ are now encoded into the output qRAM state $\sigma_j^\mathrm{out} = \sum_{i=1}^M q_{ij} |i\rangle \langle i| \otimes \rho^{\otimes k}_{j|i}$ as the routing probability. }
\label{fig:sp_circuit}
\end{figure}

With such a correlated qRAM states at hand, we can measure all the $M$ geometric relationships $\{ \Omega (\rho_{j|i}, \chi_j) \}_{i=1}^M$ in parallel by means of a SWAP test \cite{Buhrman2001Quantum}. As shown in the Fig.~\ref{fig:sp_circuit}(a), except for the input qRAM state $\sigma_j^\mathrm{in}$, we also need to prepare $\chi_j^{\otimes k}$ that is stored in the capsule register. This can be done by tracing out the index qubits of the output qRAM state $\chi^{\otimes k}_j  = \mathrm{Tr_{index}} (\sigma_j^\mathrm{out}) = \sum_{i=1}^M q_{ij} \rho^{\otimes k}_{j|i}$. Then, we introduce an ancilla qubit initialized as $|0\rangle$ and apply a Hadamard gate on it. After that a controlled $2k$-SWAP gate (controlled-$U_s$) is applied to the ancilla, capsule and bus registers. The $2k$-SWAP gate $U_s$ is defined as:
\begin{equation} \label{eqn:2kSWAP}
    U_s |\psi_0\rangle \otimes |\psi_1\rangle \otimes \dots \otimes |\psi_{2k}\rangle = |\psi_{2k}\rangle \otimes |\psi_0\rangle \otimes \dots \otimes |\psi_{2k-1}\rangle, \,
\end{equation}
which permutes $2k$ quantum states for one single cycle. After applying another Hadamard gate to the ancilla qubit and tracing out both the capsule and bus registers, we obtain the following quantum state:
\begin{equation}
    \frac{1}{2M}\sum_{i=1}^M \left[ \left(1+\mathrm{Tr}(\rho_{j|i}^k \chi_j^k)\right) |0\rangle \langle 0| \otimes |i\rangle \langle i| + \left(1-\mathrm{Tr}(\rho_{j|i}^k \chi_j^k)\right) |1\rangle \langle 1| \otimes |i\rangle \langle i| \right] \,.
\end{equation}
{ Finally, by projecting the ancilla qubit into the $|0\rangle \langle 0|$ subspace with probability $\mathcal{O}(\mathcal{A}/(2M)) \approx\mathcal{O}(1) $}, we can efficiently encode all the $M$ geometric relationships $\{ \Omega (\rho_{j|i}, \chi_j) \}_{i=1}^M$ into the index register as a diagonal quantum density matrix:
\begin{equation} \label{eqn:overlap_state}
    \omega_j = \sum_{i=1}^M \omega_{ij} |i\rangle \langle i | \,, \qquad 
    \omega_{ij} = \frac{1+\mathrm{Tr}(\rho_{j|i}^k \chi_j^k)}{\mathcal{A}} \,, \qquad 
    \mathcal{A} = \sum_{i=1}^M \left[ 1+\mathrm{Tr}(\rho_{j|i}^k \chi_j^k) \right] \, .
\end{equation}
The normalization factor $\mathcal{A}$ can be estimated in $\mathcal{O}(\log(N) / \epsilon^2)$ measurements with the error $\epsilon$ for an $N$-dimensional state.

\begin{algorithm}[t]
    \label{alg:QuantumDynamicRouting}
    \SetKwInOut{Input}{input}\SetKwInOut{Output}{output}

    \Input{the index of layer $l$, the order of moment $k$, the number of iteration $e$, the prediction quantum density matrix $\mathbf{\rho}^{l+1}_{j|i} = \mathcal{E}_{ij} (\chi^l_{i})$}
    \Output{the quantum capsule state $\chi^{l+1}_j$ in the $(l+1)$-th layer}
    \textbf{initialize}: $\forall i \in \mathcal{L}_{l}, j \in \mathcal{L}_{l+1}$: fix $\sigma^\mathrm{in}_{j} = 1/M \sum\nolimits_i \ketbra{i} \otimes (\rho^{l+1}_{j|i})^{\otimes k}$ and assign $q_{ij} \leftarrow 1/M$ to $\sigma^\mathrm{out}_{j} = \sum_i q_{ij} |i\rangle \langle i| \otimes (\rho^{l+1}_{j|i})^{\otimes k}$ \\
    \For{$e$ iterations and $\forall j \in \mathcal{L}_{l+1}$}{
        $(\chi^{l+1}_j)^{\otimes k} \leftarrow \mathrm{Tr}_\mathrm{index}(\sigma^\mathrm{out}_{j})$ \tcp*[f]{trace out the index register}\\
        $\omega_j = \sum\nolimits_i w_{ij} \ketbra{i} \leftarrow \mathrm{SWAP\_test}(\sigma^\mathrm{in}_{j}, \chi^{l+1}_j)$ \tcp*[f]{parallel update $w_{ij}$ by $\mathrm{SWAP\_test}$ in Fig.\ref{fig:sp_circuit}(a)}\\
        {$U_\omega = \exp(-\mathrm{i} t \, \omega_{j} \otimes \mathbbm{1} \otimes X) \leftarrow \mathrm{density\_matrix\_exp} (\omega_j)$} \tcp*[f]{density matrix exponentiation in Eq.\eqref{eqn:Uomega}}\\
        $U_j \leftarrow U_p U_\omega =\exp(-\mathrm{i} t \sum_i \tilde{q}_{ij} \ketbra{i} \otimes \mathbbm{1} \otimes X)$ \tcp*[f]{apply the phase-shift unitary in Eq.\eqref{eqn:Uphase}}\\
        $\sigma^\mathrm{out}_{j} \leftarrow \sum\nolimits_i q_{ij} \ketbra{i} \otimes (\rho^{l+1}_{j|i})^{\otimes k} \xleftarrow[\rm post-selection]{\rm short\, evolution} U_j (\sigma^\mathrm{in}_{j} \otimes \ketbra{0}) U_j^\dagger $ \tcp*[f]{update $\sigma^\mathrm{out}_{j}$ with new $q_{ij}$ in Fig.\ref{fig:sp_circuit}(b)}
    }
    \Return{$(\chi^{l+1}_j)^{\otimes k} = \mathrm{Tr}_\mathrm{index}(\sigma^\mathrm{out}_{j})$}
\caption{Quantum dynamic routing for QCapsNets}
\end{algorithm}

\subsection{Assign the routing coefficients}
Note that the overlap state $\omega_j$ contains the geometric information about the $j$-th capsule state $\chi_j$ with its all corresponding prediction states $\rho_{j|i}$. Now we would like to utilize $\omega_j$ to update the routing coefficients $q_{ij}$ in the output qRAM state $\sigma_j^\mathrm{out}$ in Eq.~\eqref{eqn:qRAMin}.
{ 
To begin with, we first introduce two unitary operators, the geometric unitary $U_\omega$ and the phase-bias unitary $U_p$, by preparing an extended overlap state $\tilde{\omega}_j = \omega_j \otimes \mathbbm{1} \otimes X$ and a Hamiltonian $H_p = - \sum_{i=1}^M \frac{1}{\mathcal{A}}  \ket{i} \bra{i} \otimes \mathbbm{1} \otimes X$,
\begin{align}
	U_\omega &= \exp(- \mathrm{i} \tilde{\omega}_j t ) = \exp(-\mathrm{i} t \sum_{i=1}^M \omega_{ij}  \ket{i} \bra{i} \otimes \mathbbm{1} \otimes X) \,, \label{eqn:Uomega} \\
  U_p &= \exp(- \mathrm{i} H_p t ) = \exp(+ \mathrm{i} t \sum_{i=1}^M \frac{1}{\mathcal{A}} \ket{i} \bra{i} \otimes \mathbbm{1} \otimes X) \,, \label{eqn:Uphase} 
\end{align}
where $\omega_{ij}$ is the geometric relationship defined in Eq.~\eqref{eqn:RoutingCoefficients}, $\mathbbm{1}$ is the identity operator in the bus register, and the Pauli-$X$ gate acts on another ancilla qubit. The geometric unitary $U_\omega$ can be generated by the density matrix exponentiation technique \cite{Lloyd2014Quantum}: given $\mathcal{O}(t^2/\epsilon)$ copies of the density matrix $\tilde{\omega}_j$, the unitary operation $U_\omega$ can be implemented within error $\epsilon$. The phase-bias unitary $U_p$ can also be efficiently generated by the dynamic evolution of the Hamiltonian $H_p$. Combining these two unitary operators, we generate the following unitary operator, 
\begin{align}
	U_j = U_p U_\omega 
    &= \exp[-\mathrm{i} t \sum_{i=1}^M \, \left(\frac{1+\mathrm{Tr}(\rho_{j|i}^k \chi_j^k)}{\mathcal{A}} - \frac{1}{\mathcal{A}}\right) \, \ket{i} \bra{i} \otimes \mathbbm{1} \otimes X] \nonumber \\
    &= \exp(-\mathrm{i} t \sum_{i=1}^M \, \frac{\mathrm{Tr}(\rho_{j|i}^k \chi_j^k)}{\mathcal{A}} \, \ket{i} \bra{i} \otimes \mathbbm{1} \otimes X) \equiv \exp(-\mathrm{i} t \sum_{i=1}^M \, \tilde{q}_{ij} \, \ket{i} \bra{i} \otimes \mathbbm{1} \otimes X) \nonumber \\
    &= \sum_{i=1}^M \Big[ \cos(\tilde{q}_{ij} t) \ket{i} \bra{i}\otimes \mathbbm{1} \otimes \mathbbm{1} - \mathrm{i} \sin(\tilde{q}_{ij} t) \ket{i} \bra{i} \otimes \mathbbm{1} \otimes X \Big]
    \,,
\end{align}
which encodes all the $M$ geometric relationships $\{ \Omega (\rho_{j|i}, \chi_j) \}_{i=1}^M$. We define $\tilde{q}_{ij} = \mathrm{Tr}(\rho_{j|i}^k \chi_j^k)/\mathcal{A}$ to simplify the above equation, which should not be confused with the real routing coefficients $q_{ij}$.} Accordingly, as shown in Fig.~\ref{fig:sp_circuit}(b), we are now able to assign the routing probability to each capsule by applying $U_j$ on the input qRAM state [Eq.~\eqref{eqn:qRAMin}] and the ancilla qubit:
\begin{align}
    U_j (\sigma_j^\mathrm{in} \otimes \ket{0} \bra{0}) U_j ^\dagger 
    = \frac{1}{M} \sum_{i=1}^M  \Big [ & \cos^2(\tilde{q}_{ij} t) \ket{i} \bra{i} \otimes \rho^{\otimes k}_{j|i} \otimes \ket{0} \bra{0} + \mathrm{i}\cos(\tilde{q}_{ij} t) \sin(\tilde{q}_{ij} t) \ket{i} \bra{i} \otimes \rho^{\otimes k}_{j|i} \otimes \ket{0} \bra{1} \nonumber \\
    + & \sin^2(\tilde{q}_{ij} t) \ket{i} \bra{i} \otimes \rho^{\otimes k}_{j|i} \otimes \ket{1} \bra{1} - \mathrm{i}\cos(\tilde{q}_{ij} t)\sin(\tilde{q}_{ij} t) \ket{i} \bra{i} \otimes \rho^{\otimes k}_{j|i} \otimes \ket{1} \bra{0} \Big ] \, .
\end{align}
In order to obtain a good approximation for $\sin(\tilde{q}_{ij} t) \approx \tilde{q}_{ij} t$, we set $t$ a small value and measure the ancilla qubit in $\ket{1}$ state. We remark that we can apply the amplitude amplification scheme to boost the probability of finding the state $\ket{1}$~\cite{brassard2002quantum}. In this case, the cost of measurement is around $\mathcal{O}(\sqrt{\frac{1}{\tilde{q}_{ij}^2 t^2}})\approx \mathcal{O}(\mathcal{A}/t)$. We obtain the post-selection state and assign all the $M$ geometric relationships to the output qRAM state:
\begin{equation} \label{eqn:qRAMoutUpdate}
	\sigma_j^\mathrm{out} \leftarrow \sum_{i=1}^M q_{ij} |i \rangle \langle i | \otimes \rho^{\otimes k}_{j|i}, \quad q_{ij} = \frac{|\mathrm{Tr}(\rho_{j|i}^k \chi_j^k)|^2}{\sum_{i=1}^M|\mathrm{Tr}(\rho_{j|i}^k \chi_j^k)|^2} \,.
\end{equation}
Tracing out the index qubits of the output qRAM state, the capsule state $\chi_j^{\otimes k}$ is now refined to $\mathrm{Tr_{index}} (\sigma_j^\mathrm{out}) = \sum_{i=1}^M q_{ij} \rho^{\otimes k}_{j|i}$ with new routing coefficients $q_{ij}$ containing the geometric information. Through repeating the above quantum routing procedure for few iterations (normally 3 in our numerical experiments), the routing coefficients $q_{ij}$ generally converge to constants. 

We give some additional remarks that the aforementioned oracle $O_v$ in Eq.~\eqref{eqn:qRAMin} can be treated as a generalization of the conventional qRAM from pure quantum states to mixed quantum states. A conventional qRAM is a device that, given a superposition of addresses $i$, returns a correlated set of data $D_i$: $\sum_i \alpha_i \ket{i} | \vec{0} \rangle \xrightarrow{\text{qRAM}} \sum_i \alpha_i \ket{i} | D_i \rangle$ \cite{Giovannetti2008Quantum}. On a par with the Choi-Jamiołkowski isomorphism, which establishes a correspondence between quantum channels and density matrices \cite{Choi1975Completely,Jamiolkowski1972Linear,Wilde2017Quantum}, we adopt a bijection that maps a density matrix to a vector in the computational basis $\rho_i=\sum_{m, n} (\rho_i)_{m n} \ket{m}\bra{n} \Leftrightarrow \ket{\rho_i} = \sum_{m, n} (\rho_i)_{m n} \ket{m}\ket{n}$. Hence, the conventional qRAM is able to perform $\sum_i \alpha_i \ket{i} \ket{i} \otimes (| \vec{0} \rangle | \vec{0} \rangle)^{\otimes k} \xrightarrow{\text{qRAM}} \sum_i \alpha_i \ket{i} \ket{i} \otimes [\sum_{m, n} (\rho_i)_{m n} \ket{m}\ket{n} ]^{\otimes k}$, which is equivalent to $\sum_i \alpha_i \ketbra{i} \otimes (| \vec{0} \rangle \langle \vec{0} |)^{\otimes k} \xrightarrow{O_v} \sum_i \alpha_i \ketbra{i} \otimes \rho^{\otimes k}_i$ and thus achieves the oracle $O_v$.

\subsection{Complexity Analysis}

The complexity of the quantum dynamic routing algorithm can be analyzed in the following steps. First, for a general task, the amplitude encoding method may require exponential number of operations to encode the input data. Yet, this part of the complexity can be significantly alleviated when the input data is already a quantum state, such as the classification of symmetry-protected topological phase. Second, with the help of the qRAM architecture, capsules in the same layer can be addressed in $\mathcal{O}(\log (M))$, albeit the resource for preparing the qRAM state would grow linearly with respect to $M$. Third, the geometric relationship $q_{ij}$ between capsules can be measured within $\mathcal{O}(\log (N) / \epsilon^2)$ operations via the SWAP test. We remark that although the overlap of two random quantum states decays exponentially with the number of qubits, 
only a small number of qubits inside each capsule can be sufficient for the tasks considered in our main text. In addition, the routing coefficients can be parallel assigned by $\mathcal{O}(t^2/\epsilon)$ copies of the density matrix $\tilde{\omega}_j$. In this case, the time period $t$ is set to be a small constant. The post-selection of the ancillary state requires around $\mathcal{O}(\mathcal{A}/t)\approx \mathcal{O}(M/t)$ times of measurement. The leading term of the time complexity of the dynamic routing algorithms is $\mathcal{O}(Mt\log(M)\log(N))$.

\begin{table*}[t]
    \begin{ruledtabular}
        \caption{\label{tab:notations} Summary of Notations}
        \begin{tabular}{cl} 
            Notation & Meaning \\ \hline
            $l$ & the index of layers \\
            $\mathcal{L}_l$ & the set of capsule indices in the $l$-th layer \\
            $i,j$ & the index of capsules in layers \\
            $e$ & the number of iterations in dynamic routing \\
            $u,v$ & the dimension of vectors or matrices \\
            $[H, W, C]$ & the shape of data, corresponding to the dimensions of height, width, and channel. 
            \\ \hline
            $r_{ij}$ & the routing coefficient between two capsules $i$ and $j$ \\
            $b_{ij}$ & the routing coefficient between two capsules $i$ and $j$ before \texttt{softmax}\\
            $g_{ij}$ & the geometric relationship (dot product) between two capsule vector \\
            $\mathbf{v}^l_i \in \mathbb{R}^v$ & the capsule vector of the $i$-th capsule in the $l$-th layer \\
            $\mathbf{s}^l_i \in \mathbb{R}^v$ & the capsule vector of the $i$-th capsule in the $l$-th layer before \texttt{squash}\\
            $\mathbf{u}^{l+1}_{j|i} \in \mathbb{R}^u$ & the prediction vector of the $j$-th capsule in the $(l+1)$-th layer, based on the $i$-th capsule in the $l$-th layer \\
            $\mathbf{W}_{ij} \in \mathbb{R}^{u\times v}$ & the weight matrix between two capsules $i$ and $j$ in two adjacent layers \\ \hline
            $k$ & the order of moment in quantum state overlap \\
            $\Omega$ & the geometric function that quantifies the overlaps of quantum states \\
            $\omega_{ij}$ & the geometric relationship ($k$-th moment overlap) encoded in the overlap state $\omega_j$ \\
            $q_{ij}$ & the routing coefficients between two capsule state $i$ and $j$ \\
            $\mathcal{E}_{ij}$ & the quantum channel (sub-QNN) that takes capsule state $i$ as input and prediction state $j$ as output\\
            $O_v$ & the oracle that prepares the input/output qRAM states \\
            $U_s$ & the $2k$-SWAP gate that permutes $2k$ quantum states for one single cycle \\
            $U_j$ & the unitary assigns all the geometric relationships about the capsule state $j$ to the output qRAM state \\
            $\chi_j^l \in \mathbb{C}^{v\times v}$ & the capsule state of the $i$-th capsule in the $l$-th layer \\
            $\rho_{j|i}^{l+1} \in \mathbb{C}^{u\times u}$ & the prediction state of the $j$-th capsule in the $(l+1)$-th layer, based on the $i$-th capsule in the $l$-th layer \\
            $\sigma^\mathrm{in/out}_j \in \mathbb{C}^{v\times v} \otimes \mathbb{C}^{ku\times ku}$ & the input/output qRAM states whose address registers correlate all the prediction states 
            \end{tabular}
    \end{ruledtabular}
\end{table*}

\section{Numerical details}

\label{sec:Appx_Numerics}

Here we present the technical details of our numerical simulation in Sec.~\uppercase\expandafter{\romannumeral3} A. Due to the limited computational power of the classical computer, we choose only a subset of the MNIST dataset consisting of images labeled with ``3'' and ``6'', and reduce the size of each image from $28\times28$ pixels into $16\times 16$ pixels. We normalize and encode these pixels into an $8$-qubit quantum state using the amplitude encoding method, together with an ancilla qubit.

The QCapsNet is composed of the preprocessing layer, the primary capsule layer, and the digit capsule layer. In the preprocessing layer, we apply a 5-depth PQC on the 9-qubit system. Then, we rearrange these 9 qubits into 3 capsules in the primary capsule layer as follows. These qubits are first labeled from 1 to 9 respectively, then we partial trace qubits 4-9 to compute the first capsule. Similarly, we can obtain the second capsule by tracing out qubits 1-3 and 7-9, and the third capsule by tracing out qubits 1-6.  Each capsule contains 3 qubits forming a sub-QNN $\mathcal{E}_{ij}$. We apply $\mathcal{E}_{ij}$ to capsule states $\chi^l_i$ and obtain their corresponding prediction states $\rho^{l+1}_{j|i} = \mathcal{E}_{ij} (\chi^l_i)$. In numerical simulation, capsules should be computed repeatedly. Yet for a real setup, the parameterized unitaries can be applied simultaneously without tracing out other qubits, and thus we do not need to recompute the capsule. In the quantum dynamic routing process, we fix the moment order as $k=3$, as it already results in a fast convergence.
For two-category classifications, we place two capsules in the last layer of QCapsNets. Inside each capsule $c$, we measure the average $z$-directional magnetic moment of 3 qubits, $\overline{\langle Z_c \rangle} = \sum^3_{j=1} \langle Z_j \rangle/3$. Then, the activation probability of $c$-th output capsule state $\chi_c$ is computed by the normalized amplitude $P_c = (1+\overline{\langle Z_c \rangle})/2$, which indicates the presence of label $y_c$. Consequently, we use the cross entropy as the loss function for the stand-alone classification task, 
\begin{equation}
    L_\mathrm{CE} = - \sum\nolimits_{c} y_c \text{log} P_c \, .
\end{equation}

As for the reconstruction task, the loss function has two parts. Except for the classification loss, we need to consider an extra reconstruction loss. Although we have already achieved a good performance in the classification task by use of the cross-entropy loss $L_\mathrm{CE}$, such loss function is unable to retain enough information for the output capsule state. Under this circumstance, we replace the above cross-entropy loss with a margin loss $L_\mathrm{M}$ as the first part of the loss function for the reconstruction task,
\begin{equation}
    L_\mathrm{M} = \sum\nolimits_c T_c \max (0, m^+ - P_c)^2 \nonumber + 0.5 (1-T_c) \max (0,P_c - m^-)^2 \, ,
\end{equation}
where $T_c=1$ if and only if $c$ matches the label of the input image. We set $m^+=0.9$ and $m^- = 0.1$ as constants, and define the activation probability $P_c = \mathcal{P}(\chi_c) = \text{Tr}(\chi_c^2)$ as the purity of the $c$-th output capsule state $\chi_c$. Since the margin loss $L_\mathrm{M}$ measures the purity of the capsule states, it can capture the essential quantum information that required to reconstruct the original data. 
The second part of the loss function is the reconstruction loss, where we use a mean square loss, 
\begin{equation}
	L_\mathrm{MSE} = \frac{1}{n}\sum_{i=1}^n(x_i - \hat{x}_i)^2 \,,
\end{equation}
to quantify the deviation between the reconstruction image $x$ and the original image $\hat{x}$. Here, $\hat{x}_i$ refers to the $i$-th pixel of image $\hat{x}$ and $n = 16 \times 16$ is the number of pixels. To concentrate the whole networks on the classification subroutine, we set the overall loss as $L = L_\mathrm{M} + 0.1 L_{\text{MSE}}$. Here $0.1$ is a regularization factor that scales down $L_\mathrm{MSE}$ so that it does not dominate $L_\mathrm{M}$ during training.

The code of QCapsNets is implemented by the classical version of TensorFlow \cite{Abadi2016TensorFlow}. In the optimization process, we use the automatic differentiation technique to calculate the gradient. The optimization strategy is based on the Adam optimizer. To speed up the matrix multiplication, some data is processed by the GPU version of TensorFlow. The ground state of the cluster-Ising Hamiltonian is obtained by the exact diagonalization method. {  We run our experiments on a single server with 192 processors of Intel(R) Xeon(R) Platinum 9242 CPU @ 2.30GHz. The classification task is performed on the CPU. For every single epoch, the running time ranges from $10$ minutes to $55$ minutes which depends on the structure of the capsule. The reconstruction task is achieved by an NVIDIA TITAN V GPU. The running time for each epoch is $20$ seconds.}

\section{Detailed structure of the QCapsNet}

\label{sec:Appx_Structure}

\begin{figure}[t]
    \centering
    \includegraphics[width=0.98\textwidth]{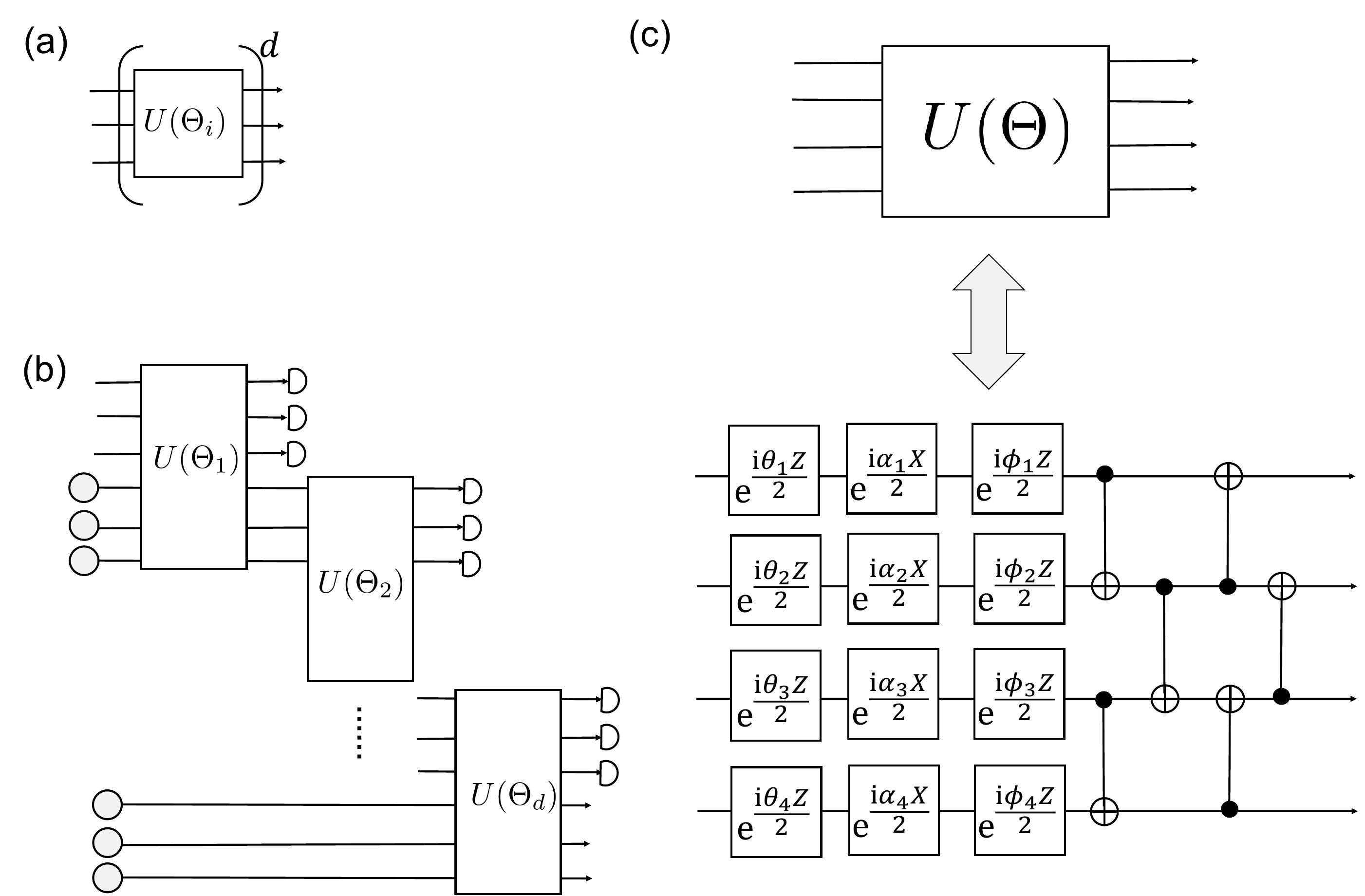}
    \caption{Detailed structures of QCapsNets.  (a) and (b) are two different sub-QNNs inside the capsule of QCapsNets. (a) is the quantum parameterized circuit (PQC). (b) is the deep quantum feedforward neural network (DQFNN).  (c) is the detailed structure of the parameterized unitary operator $U(\theta)$ with several layers. Each layer contains a set of parameterized gates for Euler rotation and CNOT gates between the nearest qubits. }
    \label{fig:structures}
\end{figure}

In our numerical simulation, we investigate two different sub-QNNs $\mathcal{E}_{ij}$ inside each capsule, namely, the parameterized quantum circuit (PQC) and the deep quantum feedforward neural network (DQFNN) . Their explicit structures are demonstrated in Fig.~\ref{fig:structures}(a,b), and the corresponding QCapsNets are dubbed PQC-Caps and DQFNN-Caps, respectively. For the PQC-Caps, we use the simple feed forward structure with a $d$-depth parameterized unitary operator $U(\theta)$. For the DQFNN-Caps, between the $l$-th layer and $(l+1)$-th layer, we first apply a parameterized quantum circuit $U(\theta)$ on the whole system and then trace out the qubits in the $l$-th layer.  The detailed structure of the parameterized quantum circuits $U(\theta)$ is shown in Fig.~\ref{fig:structures}(c), which contains several layers. Inside each layer, there is a set of parameterized gates for Euler rotation and then a sequence of nearest CNOT gates.

\bibliography{QCapsNets}
\end{document}